\documentclass[aps,prl,twocolumn,amsmath,amssymb]{revtex4}
\usepackage[english]{babel}
\usepackage{amsmath}
\usepackage{amssymb}
\usepackage{graphicx}
\usepackage{color}

\begin{document}

\title{Two-fluid model of a Bose-Einstein condensate in the cavity optomechanical regime}

\author{Daniel S. Goldbaum$^{1}$}
\author{Keye Zhang$^{2}$}
\author{Pierre Meystre$^{1}$}
\date{\today}

\affiliation{\begin{tabular}{c}
$^{1}$B2 Institute, Dept. of Physics and College of Optical Sciences, The
University of Arizona, Tucson, AZ 85721, USA \\
$^{2}$State Key Laboratory of Precision Spectroscopy, Dept. of Physics, East
China Normal University, Shanghai 200062, China
\end{tabular}}

\begin{abstract}
We analyze an atomic Bose-Einstein condensate trapped in a high-$Q$ optical cavity driven by a feeble optical field, a situation formally analogous to the central paradigm of cavity optomechanics [Brennecke {\it et al}. (Science, \textbf{322}, 235 (2008)]. We account for two-body interactions via a two-fluid model that retains the intuitive appeal of the optomechanical two-mode description. The Bogoliubov excitation spectrum of this system comprises a gapped upper branch and a lower branch that can include an unstable excitation mode.
\end{abstract}

\maketitle

\newcommand{\Psidx}{\hat{\Psi}^{\dagger} \negthinspace \left( x \right)}
\newcommand{\Psix}{\hat{\Psi} \negthinspace \left( x \right)}
\newcommand{\phiqx}{\phi_q \negthinspace \left( x \right)}
\newcommand{\phiqpx}{\phi_{q^{\prime}} \negthinspace \left( x \right)}
\newcommand{\phisqx}{\phi_q^* \negthinspace \left( x \right)}
\newcommand{\phisqpx}{\phi_{q^{\prime}}^* \negthinspace \left( x \right)}
\newcommand{\bq}{\hat{b}_q}
\newcommand{\bdq}{\hat{b}_q^{\dagger}}
\newcommand{\epsqo}{\epsilon_q^0}
\newcommand{\bdqpr}{\hat{b}_{q+r}^{\dagger}}
\newcommand{\bdqpmr}{\hat{b}_{q^{\prime}-r}^{\dagger}}
\newcommand{\bqp}{\hat{b}_{q^{\prime}}}
\newcommand{\bdqp}{\hat{b}_{q^{\prime}}^{\dagger}}
\newcommand{\qp}{q^{\prime}}
\newcommand{\bnq}{\hat{b}_{-q}}
\newcommand{\bdnq}{\hat{b}_{-q}^{\dagger}}
\newcommand{\vq}{v_q}
\newcommand{\uq}{u_q}
\newcommand{\btq}{\hat{\beta}_q}
\newcommand{\btdnq}{\hat{\beta}_{-q}^{\dagger}}
\newcommand{\epsq}{\epsilon_q}
\newcommand{\btdq}{\hat{\beta}_{q}^{\dagger}}
\newcommand{\etr}{\tilde{\epsilon}_r}
\newcommand{\ah}{\hat{a}}
\newcommand{\ahd}{\hat{a}^{\dagger}}
\newcommand{\lamqx}{\lambda_{q} \negthinspace \left( x \right)}
\newcommand{\lamsqx}{\lambda_{q}^* \negthinspace \left( x \right)}
\newcommand{\lamsqpx}{\lambda_{q^{\prime}}^* \negthinspace \left( x \right)}
\newcommand{\gq}{\hat{\gamma}_q}
\newcommand{\gdq}{\hat{\gamma}_q^{\dagger}}
\newcommand{\gdqp}{\hat{\gamma}_{q^{\prime}}^{\dagger}}
\newcommand{\gqp}{\hat{\gamma}_{q^{\prime}}}
\newcommand{\bqtk}{\hat{b}_{q+2k}}
\newcommand{\bqmtk}{\hat{b}_{q-2k}}
\newcommand{\bdqtk}{\hat{b}_{q+2k}^{\dagger}}
\newcommand{\bdqmtk}{\hat{b}_{q-2k}^{\dagger}}
\newcommand{\Hod}{\hat{\mathcal{H}}_{0,\text{d}}}
\newcommand{\Hogo}{\hat{\mathcal{H}}_{0,\gamma \Omega}}
\newcommand{\Hobg}{\hat{\mathcal{H}}_{0,b \gamma}}
\newcommand{\Ho}{\hat{\mathcal{H}}_0}
\newcommand{\lp}{\left(}
\newcommand{\rp}{\right)}
\newcommand{\lb}{\left[}
\newcommand{\rb}{\right]}
\newcommand{\epstko}{\epsilon_{2k}^0}
\newcommand{\Hap}{\hat{\mathcal{H}}_{\text{ap} }}
\newcommand{\ahatd}{\hat{a}^{\dagger}}
\newcommand{\ahat}{\hat{a}}
\newcommand{\intol}{\int_0^{L} dx \,}
\newcommand{\intll}{\int_{-L/2}^{L/2} dx \,}
\newcommand{\Haa}{\hat{\mathcal{H}}_{\text{aa}}}
\newcommand{\Haao}{\hat{\mathcal{H}}_{AA,0}}
\newcommand{\Haaone}{\hat{\mathcal{H}}_{AA,1}}
\newcommand{\Haat}{\hat{\mathcal{H}}_{AA,2}}
\newcommand{\bmq}{\hat{b}_{-q}}
\newcommand{\bdmq}{\hat{b}_{-q}^{\dagger}}
\newcommand{\gdnq}{\hat{\gamma}_{-q}^{\dagger}}
\newcommand{\gnq}{\hat{\gamma}_{-q}}
\newcommand{\go}{\hat{\gamma}_0}
\newcommand{\gdo}{\hat{\gamma}_0^{\dagger}}
\newcommand{\Hcal}{\hat{\mathcal{H}}}
\newcommand{\dbq}{\dot{\hat{b}}_q}
\newcommand{\dbdnq}{\dot{\hat{b}}_{-q}^{\dagger}}
\newcommand{\dgq}{\dot{\hat{\gamma}}_q}
\newcommand{\dgdnq}{\dot{\hat{\gamma}}_{-q}^{\dagger}}
\newcommand{\dgo}{\dot{\hat{\gamma}}_0}
\newcommand{\dgdo}{\dot{\hat{\gamma}}_0^{\dagger}}
\newcommand{\dahat}{\dot{\hat{a}}}
\newcommand{\epsrt}{\epsilon \negthinspace \left(r, \tau \right)}
\newcommand{\epstk}{\epsilon_{2k} \negthinspace \lp r, \tau  \rp }
\newcommand{\Hk}{\hat{\mathcal{H}}_{\text{kin}}}
\newcommand{\kronqpq}{\delta_{q^{\prime} q}}
\newcommand{\Erec}{E_{\text{rec}}}
\newcommand{\epsbq}{\epsilon_{\beta} \negthinspace \lp q \rp}
\newcommand{\epsaq}{\epsilon_{\alpha} \negthinspace \lp q \rp}
\newcommand{\betq}{\hat{\beta}_q}
\newcommand{\alfq}{\hat{\alpha}_q}
\newcommand{\betdq}{\hat{\beta}^{\dagger}_q}
\newcommand{\alfdq}{\hat{\alpha}^{\dagger}_q}
\newcommand{\epsao}{\epsilon_{\alpha}^{o}}
\newcommand{\epsaot}{\epsilon_{\alpha}^{o \, 2}}
\newcommand{\Aqx}{A_q \negthinspace \lp x \rp}
\newcommand{\Bqx}{B_q \negthinspace \lp x \rp}
\newcommand{\Bnox}{B_0 \negthinspace \lp x \rp}
\newcommand{\Aox}{A_0 \negthinspace \lp x \rp}
\newcommand{\alfo}{\hat{\alpha}_0}
\newcommand{\alfdo}{\hat{\alpha}^{\dagger}_0}
\newcommand{\epsaof}{\epsilon_{\alpha}^{o \, 4}}
\newcommand{\Hq}{\hat{\mathcal{H}}_q}
\newcommand{\Hgs}{\hat{\mathcal{H}}_{\text gs}}
\newcommand{\epsbo}{\epsilon_{\beta}^{o}}
\newcommand{\alfnq}{\hat{\alpha}_{-q}}
\newcommand{\alfdnq}{\hat{\alpha}^{\dagger}_{-q}}
\newcommand{\betnq}{\beta_{-q}}
\newcommand{\betdnq}{\beta^{\dagger}_{-q}}
\newcommand{\Hni}{\hat{\mathcal{H}}_{\text{ni}}}
\newcommand{\bo}{\hat{b}_0}
\newcommand{\btk}{\hat{b}_{2k}}
\newcommand{\bntk}{\hat{b}_{-2k}}
\newcommand{\On}{\hat{\Omega}_{0}}
\newcommand{\btoo}{\hat{b}_2}
\newcommand{\bntoo}{\hat{b}_{-2}}
\newcommand{\bqt}{\hat{b}_{q+2}}
\newcommand{\bqmt}{\hat{b}_{q-2}}
\newcommand{\Omq}{\hat{\Omega}_{q}}
\newcommand{\Omdq}{\hat{\Omega}^{\dagger}_{q}}
\newcommand{\Hnid}{\hat{\mathcal{H}}_{\text{ni,d}}}
\newcommand{\Hnigo}{\hat{\mathcal{H}}_{\text{ni,$\gamma \Omega$}}}
\newcommand{\Hnibg}{\hat{\mathcal{H}}_{\text{ni,$b \gamma$ }}}
\newcommand{\co}{{\hat c}_0}
\newcommand{\so}{{\hat s}_0}

Cavity optomechanics offers much promise for applications ranging from the detection of weak forces and small displacements to fundamental studies of the transition between the quantum and the classical world~\cite{Kippenberg:07}. One major stepping stone toward these objectives is to optically cool one or a few modes of vibration of a moving mirror to near its quantum-mechanical ground state. This goal is likely to be achieved in the near future, in situations ranging from micro- and nano-mechanical systems~\cite{LaHaye04022004,Kleckner:SubkelvinCooling,Metzger:CavityCoolingMicrolever,ekinci:2682,Schliesser:2009fj,Gigan_Nature_444,Thompson:2008xy} to large mirrors as used in gravitational wave detectors~\cite{corbitt:150802,anderlini:013001,corbitt:160801, LIGO2009}.

In parallel to these developments, recent studies of ultracold atomic gases trapped in optical cavities have provided an alternative path to the study of cavity optomechanics~\cite{Gupta:PRL_99_212601, Murch:Nature_4_561, Brennecke10102008}. Here the excitation of a collective mode of the cold gas plays the role of the vibrational mode of the mirror. Optomechanical studies using cold gases are appealing for a number of reasons. Most importantly perhaps, methods for preparing the atoms in their motional ground state are well developed, allowing one to prepare the effective oscillator mode in its quantum ground state. Also, due to the high cooperativity of the cold atomic gas, one reaches the strong-coupling limit of cavity quantum-electrodynamics (QED) for mean intracavity photon numbers on the order of unity, offering a promising avenue to combined studies of cavity QED and cavity optomechanics.

The optomechanical behavior of ultracold atomic gases was demonstrated in experiments by Gupta {\it et al}.~\cite{Gupta:PRL_99_212601} and by Murch {\it et al}.~\cite{Murch:Nature_4_561}, who showed that the cavity field coupling to a collective center-of-mass excitation of the atoms results in oscillatory displacement of the gas, as well as by Brennecke {\it et al}.~\cite{Brennecke10102008}, who studied the coupling between a density modulated Bose-Einstein condensate (BEC) and the cavity field. This group also developed a simple two-mode description of the BEC that clearly illustrates the formal analogy with the moving mirror system, and is the basis for an accurate description of the observed dynamics.

This Letter extends that model to calculate the low energy excitation spectrum of a BEC coupled to a weak light field. One key element is to describe the condensate by a two-{\it fluid} model that accounts for mode-broadening due to two-body collisions, while retaining the intuitive appeal of the two-mode description. We calculate the two-branch excitation spectrum via a Bogoliubov-type perturbative treatment of the collisions~\cite{PethickandSmith, Bogoliubov}. The upper branch of the spectrum is gapped, while the lower branch can become positive imaginary for small momenta, corresponding to an unstable excitation mode that grows exponentially in time.

We consider a BEC trapped inside a Fabry-P{\' e}rot cavity and interacting with a standing wave light-field generated by a pump laser of wave number $k$ parallel to the cavity axis. For simplicity, we restrict our calculation to one spatial dimension along the propagation direction of the pump laser, assume that the trap is soft enough so that its impact on the ground state of the BEC can be ignored, and work at $T=0$. In this Letter we focus on the effects of both atom-light and atom-atom interactions on the static properties of the condensate.

This system is described by the Hamiltonian $ \Hcal = \Hk + \Hap + \Haa $, where $\Hk$ is the kinetic energy term, and
\begin{equation}
\Hap = U_{0} \, \ahatd \ahat \intll \Psidx \left[  \cos^2 \negthinspace \lp k x \rp \right] \Psix
\label{A}
\end{equation}
is the atom-field interaction, where $L$ is the cavity length, $\hbar=1$, and $U_0$ the single-photon light shift of the atoms. Finally, $\Haa$ describes two-body collisions in the familiar fashion, with $U_{aa}$ the atom-atom interaction strength.

We proceed by expanding the Schr{\" o}dinger field operator of the condensate in terms of plane waves, \begin{equation}
\Psix = \sum_{q} \phi_q \negthinspace \lp x \rp \bq \, ,
\label{A2}
\end{equation}
where $\phi_q \negthinspace \lp x \rp = \frac{1}{\sqrt{L}} e^{i q x}$, $\bq$ is the corresponding boson annihilation operator, and the sum runs over all momenta $q$ satisfying periodic boundary conditions. In general, the atom-field interaction results in the recoil of atoms by $\pm2 \ell k$, where $\ell$ is an integer. Typically many momentum side modes are excited, but for feeble fields we expect intuitively that only the $\pm 2k$ side modes are appreciably occupied. The field operator then reduces to
\begin{equation}
\Psix = \phi_0 \negthinspace \lp x \rp \hat{b}_0 + \phi_{2k} \negthinspace \lp x \rp \btk + \phi_{-2k} \negthinspace \lp x \rp \bntk \, .
\label{B}
\end{equation}
It is this truncation that results in the analogy between the optically driven BEC and radiation pressure driven mirrors.

The validity of the restricted plane wave basis is determined by the ratio between the effective Rabi frequency $U_0 n_c$ and the recoil energy $\Erec= k^2/2 m$, where $m$ is the atomic mass. One can get a sense of its region of validity by numerically calculating the mean-field ground state of the BEC subject to the potential $V \negthinspace \lp x \rp = U_0 n_c \cos^2 \negthinspace \lp kx \rp$, where $n_c$ is the mean intracavity photon number, and evaluating the occupation probability $P_{\text{ho}}$ of states with $\ell > 1$. When $P_{\text{ho}} \ll 1$, the truncation~(\ref{B}) captures the qualitative features of the system. Using the parameters of Ref.~\cite{Brennecke10102008}~\footnote{Experimental parameters for $^{87}$ Rb are: two-body collision strength $g \sim 5.6 \times 10^{-51} \, \text{J} \cdot \text{m}^{3}$, atomic density $n=3 \times 10^{20} \, \text{m}^{-3}$, laser wavelength $\lambda_L = 780 \, \text{nm}$, single-photon light shift $U_0 = 2 \pi \times 3.7 \, \text{kHz}$}, this calculation suggests that we are safely in that regime for $n_c \lesssim 25 $ ($P_{\rm ho} \simeq 0.08$). Since the optical field couples the condensate to the $q=\pm 2k$ side modes symmetrically we expand $\Psix$ with respect to symmetric (`cosine') and anti-symmetric (`sine') operators,
\begin{equation}
\Psix = \phi_0 \negthinspace \lp x \rp \bo + \sqrt{2/L} \cos \negthinspace \lp 2 k x \rp \hat{c}_0 + i \sqrt{2/L} \sin \negthinspace \lp 2 k x \rp \hat{s}_0,
\label{B2}
\end{equation}
where $\co = \frac{1}{\sqrt{2}} \lp \btk + \bntk \rp$ and $\so = \frac{1}{\sqrt{2}} \lp \btk - \bntk \rp$ obey boson commutation relations. In the absence of optical fields and in the collisionless regime the many-atom ground state is a pure condensate where each atom occupies the $\bo$ mode. Switching on the feeble optical field couples the mode $\bo$ to ${\hat c}_0$, but leaves ${\hat s}_0$ unoccupied. This is the situation described by the two-mode (or single-mirror) optomechanical model of Ref.~\cite{Brennecke10102008}.\footnote{This model can easily be extended to situations where higher-order momentum modes are excited by photon recoil, generalizing Eq.~(\ref{sinecosine}) to include these modes. The situation becomes then analogous to an optomechanical system with $\ell$ mirrors coupled by the intracavity optical field.}

Two-body collisions complicate the situation by scattering atoms from the modes $q=\left\{0,\pm2k \right\}$ into nearby momentum states. The momentum distribution then consists of three sub-distributions, each centered about one of $q=\left\{0,\pm2k \right\}$. If these sub-distributions are narrow enough that they do not appreciably overlap, then one may treat each of them as a distinct Bose gas. The symmetry of the atom-field interaction suggests again the introduction of `cosine' and 'sine' operators
\newcommand{\cq}{{\hat c}_q}
\newcommand{\sq}{{\hat s}_q}
\begin{eqnarray}
\cq &=& \frac{1}{\sqrt{2}} \lp \bqtk + \bqmtk \rp, \nonumber \\
\sq &=& \frac{1}{\sqrt{2}} \lp \bqtk - \bqmtk \rp,
\label{sinecosine}
\end{eqnarray}
where the condition $\lvert q \rvert < k$ ensures that we can treat the components of the BEC described by the modes $ \left\{ \bq, \sq, \cq  \right\}$, as three distinct fluids. The widths of their momentum distributions depend only on the mean-field interaction energy $g n$, where  $g=\lp 4\pi \hbar^2 a_s  \rp/m$ is the two-body interaction strength. The other relevant energy scale is $\Erec$, hence one can get a sense for the validity of the three-fluid model from the Bogoliubov ground state occupation distribution as a function of the ratio $r=g n/\Erec$.  For the conditions of Ref.~\cite{Brennecke10102008} we have $r \simeq 0.67$ and we find numerically that the three-fluid model is indeed valid.

In the case where the intracavity field is in a Fock state $|n_c\rangle$, or for  classical fields, we have that $a \rightarrow \sqrt{n_c}$, with $n_c$ the mean intracavity photon number.~\footnote{For arbitrary quantized fields one can diagonalize the collisionless Hamiltonian in analogy with the familiar dressed states approach of quantum optics.} Within the three-fluid description, the collisionless Hamiltonian becomes, in dimensionless units where lengths are in units of $k^{-1}$ and energies are in units of $\Erec$,
\begin{eqnarray}
\Ho &=&  \sqrt{2} \varepsilon \hat{N}
+ \sum_q \lb q^2 \,  \bdq \bq + \lp q^2 + 4 \rp \lp \cq^\dagger \cq + \sq^\dagger \sq \rp \rb \nonumber \\
&+& \varepsilon \sum_q \lb {\hat c}_q^\dagger \bq + \bdq {\hat c}_q   \rb
+ 4\sum_q  q \,  \lp {\hat s}_q^\dagger {\hat c}_q  + {\hat c}_q^\dagger {\hat s}_q \rp.
\label{h0_unitless}
\end{eqnarray}
Here $\hat{N}$ is the total particle number operator,
\begin{equation}
\varepsilon =\lp \hbar U_0 n_c \rp/\lp2 \sqrt{2} \Erec \rp,
\end{equation}
and the sums are understood to span $-1 < q < 1$.

The first term in $\Ho$ is an unimportant energy shift, the second term is the kinetic energy, the third term describes the optical coupling of the modes centered around $q=0$ $( \bq )$ to the `cosine'-modes, and the fourth term accounts for the kinetic coupling of the `sine' and `cosine' modes. The effect of collisions is implicit in the inclusion of $q\neq 0$ terms.

The reduction from a three-fluid to a two-fluid description proceeds by neglecting the last term in the Hamiltonian~\eqref{h0_unitless}. The validity of this step depends on the ratio between the couplings in equation~\eqref{h0_unitless}.  Specifically, for $4 q/ \varepsilon \ll 1$, the true ground state has all qualitative features of the two-fluid ground state, the sine mode contributing only a small perturbation. The two-fluid approximation is particularly good for low momenta and $n_c$ relatively large, but still sufficiently small that the population of higher momentum side modes remains negligible. Its validity can also be readily judged graphically, by comparing the collisionless spectra of the two-fluid approximation of $\Ho$ to that of equation~\eqref{h0_unitless}.

We  mentioned already that the Bogoliubov spectrum of the two-fluid system consists of two branches. We concentrate in the following mostly on the lower branch of the spectrum. Carrying out an analysis using the experimental parameters of Brennecke {\it et al}.~\cite{Brennecke10102008}, and $n_c = 10$ we found that the inclusion of the `sine' mode results in a negligible energy difference for $q<0.1$, while $E_{\text{2F}}-E_{\text{3F}} \simeq 0.2 \Erec$ at $q=0.5$. A related test is to calculate the occupation probability of the sine-mode in the lower branch of the three-fluid model. This probability is virtually zero for $q < 0.2$, and is $\sim 0.025$ at $q=0.5$.

\begin{figure}
\includegraphics[width=11 cm]{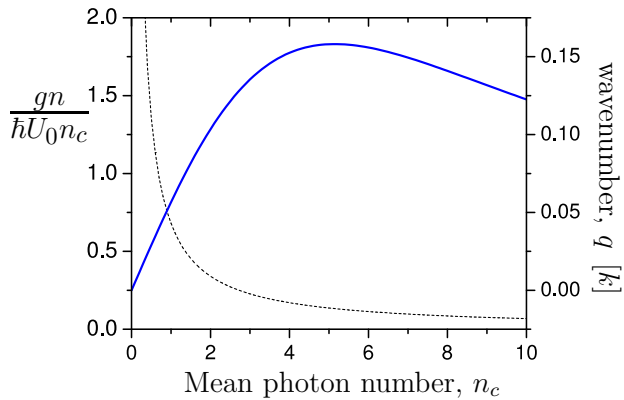}
\caption{Dotted line: $g n/\hbar U_0 n_c$, as a function of $n_c$. Solid line: Unstable region of the two-fluid system -- the lower branch of the Bogoliubov spectrum is unstable for values of $q$ below that line. In the region where a perturbative treatment of the collisions is valid, the size of the unstable region shrinks as $n_c$ increases. Parameters as in Ref.~\cite{Brennecke10102008} (see also endnote).  }
\label{perturb}
\end{figure}

The two-fluid version of the Hamiltonian~\eqref{h0_unitless} is readily diagonalized in terms of linear combinations of $\bq$ and $\cq$. Carrying out straightforward algebra we find
\begin{equation}
\Ho = \sqrt{2} \varepsilon \hat{N} + \sum_q \lb \epsilon_\beta \lp q \rp \betdq \betq +\epsilon_\gamma \lp q \rp {\hat \gamma}_q^\dagger {\hat \gamma}_q  \rb \, ,
\end{equation}
where
\begin{equation}
\epsilon_\beta \lp q \rp = 2 - \sqrt{4+\varepsilon^2} + q^2; \,\,\,\,
\epsilon_\gamma \lp q \rp = 2 + \sqrt{4 + \varepsilon^2} + q^2.
\end{equation}
The bosonic ``dressed basis'' operators are related to the bare operators by \begin{equation}
\lp \begin{array}{c} \betq  \\ {\hat \gamma}_q   \end{array} \rp =
\lp \begin{array}{cc} \mu & -\nu \\ \nu & \mu \end{array} \rp
\lp \begin{array}{c} \bq  \\ {\hat c}_q   \end{array} \rp
\end{equation}
where $\mu= \epsilon_\gamma^0/\sqrt{\epsilon_{\gamma}^{0 \, 2} + \varepsilon^2}$, $\nu=y/\sqrt{ \epsilon_{\gamma}^{0 \, 2} + \varepsilon^2}$, and $\epsilon_\gamma^0 = \epsilon_\gamma(0)$.

A Bogoliubov-type perturbative treatment of two-body collisions is appropriate provided that $g n$ is small compared to $\hbar U_0 n_c$. Figure~\ref{perturb} plots $\lp g n \rp/\lp \hbar U_0 n_c \rp$ as a function of $n_c$, and indicates that already for very modest values of $n_c$ the optical interaction dominates over the collisions. In this regime, we apply the Bogoliubov approach to the ground dressed state, that is, $\hat{\beta}_0 \rightarrow \sqrt{N_0}$, where $N_0$ is the corresponding macroscopic population. Then we keep only terms of quadratic order or less in the operators ${\hat \beta}_q$ and ${\hat \gamma}_q$, and follow the usual procedure~\cite{PethickandSmith} to arrive at a linear eigenvalue equation for the Bogoliubov excitation spectrum.

Labeling the diagonal operator for this approximation of $\Hcal$ by \begin{equation}
\hat{\Gamma}_q = u_q \betq + v_q \betdnq + w_q {\hat \gamma}_q + z_q {\hat \gamma}^{\dagger}_{-q},
\end{equation}
we find that the corresponding eigenvalue equation is
\begin{equation}
\lp \mathbf{M} - \epsilon_q \mathbf{I} \rp \mathbf{V}_q =0,
\end{equation}
where $\mathbf{V}_q= \lb u_q \, v_q \, w_q \, z_q \rb^{T}$, $\mathbf{I}$ is
the identity matrix, and
\begin{equation}
\mathbf{M}=\lb \begin{array}{cccc} K_1 & J_1 r & 2 J_2 r &  J_2 r \\
- J_1 r& -K_1 & - J_2 r & -2 J_2 r\\
2 J_2 r & J_2 r & K_2 & J_3 r \\
-J_2 r & -2 J_2 r& -J_3 r& -K_2
\end{array} \rb \, .
\end{equation}
The matrix elements of $\mathbf{M}$ are $K_1 = J_1 r + q^2$, $K_2=  \lp 2 J_3-J_1 \rp r+ 2 \sqrt{4 + y^2} +q^2$, $J_1= \frac{ 1 }{2} \lp 3 \nu^4 + 12 \nu^2 \mu^2 + 2 \mu^4 \rp$, $J_2= \frac{ 1 }{2} \lp 3 \nu^3 \mu -4 \nu \mu^3 \rp$, and $J_3= \frac{ 1 }{2} \lp 2 \nu^4 -3 \nu^2 \mu^2 + 2 \mu^4 \rp$, indicating that $\mathbf{M}$ is characterized by the three coupling constants $ \left\{ J_1, J_2, J_3 \right\}$. $J_1 g n$ and $J_3 g n$ can be thought of as the mean-field interaction energy per-particle corresponding to two-body collisions of atoms initially occupying the lower and upper branch of the dressed spectrum, respectively, and $J_2 g n$ sets the per-particle energy scale for the interaction between the two fluids. For $n_c>0$, $J_2$ is always negative, indicative of an attractive interaction.

\begin{figure}
\includegraphics[width=8 cm]{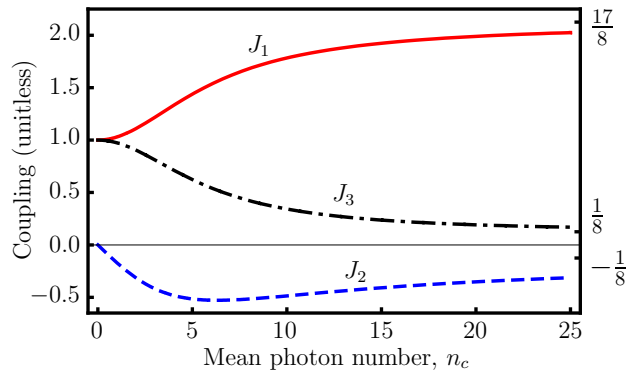}
\caption{Coupling parameters $J_1$, $J_2$, and $J_3$ as a function of $n_c$. The labeled ticks on the right-hand side of the frame denote the J-parameter values in the limit $n_c \rightarrow \infty$. Parameters as in Fig.~1.}
\label{J}
\end{figure}

Figure~\ref{J} shows these coupling constants as a function of $n_c$. For $n_c \rightarrow 0$ one recovers the familiar Bogoliubov spectrum of a weakly interacting Bose gas. For $n_c \rightarrow \infty$, $J_1=\frac{17}{8}$, $J_2=-\frac{1}{8}$ and $J_3=\frac{1}{8}$ and the lower branch of the excitation spectrum assumes the traditional Bogoliubov form characterized by the mean-field interaction energy per particle $\frac{17}{8} g n$ --- however, as already discussed, the two-fluid model breaks down in that limit. Still, Fig.~\ref{J} shows that the J-parameters approach their limiting values even for photon numbers well within the optomechanical regime, which demonstrates the relevance of the $n_c \rightarrow \infty$ limit.

\begin{figure}
\includegraphics[width=8.5 cm]{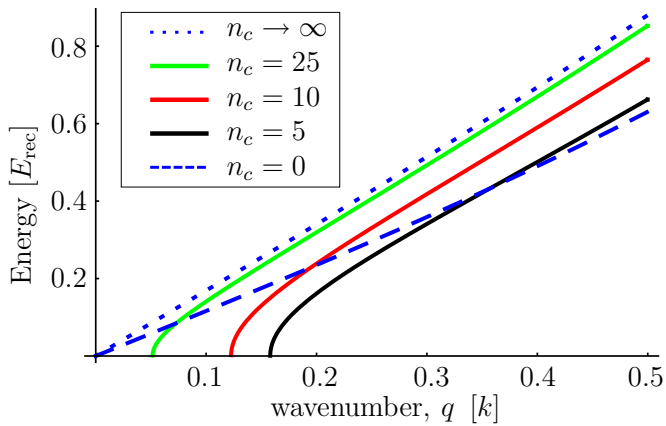}
\caption{Lower branch of the Bogoliubov spectrum of the two-fluid system. For small values of $q$ it is positive imaginary, corresponding to an excitation mode that grows exponentially in time. The instability disappears for $n_c \rightarrow 0$ $\lp n_c \rightarrow \infty \rp$, in which case we recover the conventional spectrum of a scalar condensate characterized by the mean interaction energy $gn$ ($17 gn / 8$). Parameters as in Fig. 1, and $\Erec=\hbar^2 k^2/2m$ is the recoil energy.}
\label{bogol}
\end{figure}

The upper branch of the spectrum corresponds to the dressed mode whose main contribution stems from the recoiled component of the condensate (the `cosine mode'), while the lower branch has a similar correspondence to the `condensate mode'. For vanishingly small intracavity fields, the upper branch is separated from the lower branch by $4 E_{\rm rec}$, as expected. More interesting is the lower branch displayed in Fig.~\ref{bogol}, which is characterized by a region of small momenta where the spectrum becomes positive imaginary, indicative of an instability that grows exponentially in time. This instability finds its origin in the attractive interaction between the two fluids, and its existence means that the two-fluid ground state is not the true ground state. For larger values of $q$, the lower-branch spectrum resembles a conventional Bogoliubov spectrum, and as $n_c$ increases, varies continuously between the limits $n_c=0$ (corresponding to interaction strength $g n$), and $n_c \rightarrow \infty$ (corresponding to $17 g n/8$).

The size of the unstable region depends on the strength of the atom-field interaction relative to the two-body scattering. Figure~\ref{perturb} shows the size of the unstable domain as a function of $n_c$. Below the solid line the lower branch of the Bogoliubov excitation spectrum is purely imaginary with positive magnitude. Figure~\ref{perturb} shows that this region grows in the limit where perturbation theory ceases to be valid. For $n_c \gtrsim 5$, the unstable region shrinks as $n_c$ increases, and vanishes in the limit $\frac{g n}{\hbar U_0 n_c} \rightarrow \infty$.

One important observation is that the Taylor series with respect to $q \, (>0)$ for the lower branch of the spectrum only includes terms of even parity. In contrast, the corresponding Taylor series for the conventional Bogoliubov spectrum has only odd terms. As is shown in Fig.~\ref{bogol}, this distinction is most important for small $q$ where the lowest order terms dominate, and the conventional Bogoliubov spectrum is approximately linear.

In summary we have investigated the low energy excitation spectrum of a BEC in the ``optomechanical region'' where there is a clear analogy between the dynamics of the collective excitations of the BEC, and the motion of a radiation driven moving mirror. We have extended the two-mode model that describes a collisionless BEC to a two-fluid model that accounts for two-body collisions, and have determined their effect perturbatively. We found that the atom-atom interactions are dressed by the light-field -- the most striking effect being the effective attraction between the different fluids. This attraction can lead to an unstable ground state characterized by exponentially growing excitations at small momenta. At larger momenta the low energy spectrum resembles a conventional Bogoliubov spectrum with enhanced atom-atom interactions. Further work will include an analysis of a ``multiple mirror'' situation, as well as a full quantum description of the optical field.

DSG acknowledges stimulating conversations with Julia Meyer.
This work is supported in part by the US Office of Naval Research, by the National Science Foundation, and by the US Army Research Office.

\end{document}